\documentclass[
	letterpaper, 
	10pt, 
	twoside, 
]{LTJournalArticle}

\runninghead{QNMs of Near-Extremal Black Branes} 

\footertext{} 

\title{Quasinormal Modes of Near-Extremal\\Electric and Magnetic Black Branes} 

\author{%
	Swapnil Nitin Shah\thanks{Electronic address: \href{mailto:snshah27@uw.edu}{snshah27@uw.edu}}\\
}

\date{\small{Department of Physics, University of Washington}}

\renewcommand{\maketitlehookd}{%
	\begin{abstract}
		\noindent Gauge-gravity duality provides a robust mathematical framework for studying the behavior of strongly coupled non-abelian plasmas both near and far away from thermodynamic equilibrium. In particular, their near-equilibrium transport coefficients such as viscosity, conductivity, diffusion constants, etc. can be determined from poles of the retarded Green's function which are the dissipative eigenmodes i.e., the quasinormal modes (QNMs) of the dual gravitational field equations. The AdS\textsubscript{5}/CFT\textsubscript{4} correspondence admits the description of a strongly coupled \( \mathcal{N} \)= 4 Supersymmetric Yang Mills (SYM) plasma at non-zero temperature as a dual AdS\textsubscript{5} black brane geometry. We demonstrate the application of pseudospectral methods to solving the dual Einstein field equations using the example of homogenous isotropization in \( \mathcal{N} \)= 4 SYM plasma far from equilibrium. Using this framework, we also compute the quasinormal modes of electrically (Reissner-Nordstrom) and magnetically charged AdS\textsubscript{5} black branes for the case of vanishing spatial momenta. The near-extremal behavior of these QNMs is analyzed for both types of black branes. 
	\end{abstract}
}

\begin{document}

\maketitle 

\pagebreak

\setcounter{tocdepth}{2}
\tableofcontents

\section{Introduction}

\subsection{Quark Gluon Plasma}

Quark gluon plasma (QGP) is a state of matter in which quarks and gluons comprising hadrons become free from their bound states at high energy densities. Present understanding of cosmology suggests that all hadronic matter from 10\textsuperscript{-12} to 10\textsuperscript{-6} seconds after the Big Bang existed in the form of QGP. This state of matter was first observed in heavy ion collision experiments conducted at Brookhaven National Laboratory (BNL) in 2005. Quantum Chromodynamics (QCD) is the fundamental theory of strong interactions between quarks and gluons. Asymptotic freedom in QCD states that the interaction between quarks and gluons gets increasingly weak at extremely high energies where perturbative treatment is applicable. This approach, however, fails in describing dynamics of QGP where the coupling between quarks and gluons is strong. Various attempts using non-perturbative techniques such as lattice QCD have been made to study properties of QGP. One of the important predictions of lattice QCD is the crossover temperature $\sim$200MeV from the confined hadronic phase to the deconfined QGP phase as illustrated in the hypothesized QCD phase diagram (Fig. 1)\autocite{Ar:2014}. These methods, however, run into problems with computing real time correlation functions at strong coupling and non-zero temperature, which are important for describing non-equilibrium processes. 

\begin{figure} 
	\centering
	\includegraphics[width=0.5\textwidth]{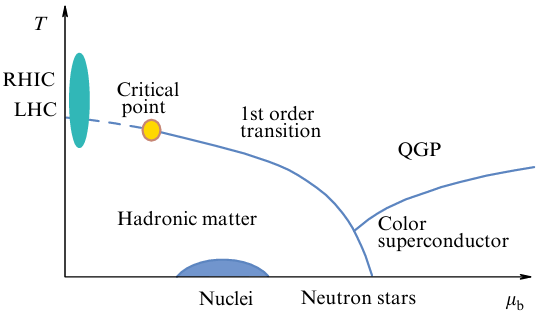}
	\caption{QCD phase diagram. Source: \textcite{Ar:2014}}
	\label{fig:Fig1}
\end{figure}

In this deconfined QGP phase, the colored partons can propagate over large distances which accounts for non-trivial collective behavior. Consider a non-central collision of two Lorentz contracted nuclei in the plane normal to the collision direction (Fig. 2) \autocite{Ar:2014}. The two nuclei only interact in the 'almond' shaped region and the parts outside do not interact. If the observed hadron jets emanating from the collision were assumed to form from individual nucleon-nucleon collisions in the almond shaped region, the resulting hadron jets would be uniformly distributed over the azimuthal angle in the collision plane, independent of the shape of the interaction region. This is in conflict with experimental data obtained at Relativistic Heavy Ion Collider (RHIC) and Large Hadron Collider (LHC). Only strong interactions between partons in a fluid medium with low shear viscosity can prevent the loss of asymmetry information from the collision which has been experimentally verified. This yields a straightforward treatment of the QGP medium near thermal equilibrium in relativistic hydrodynamics.

\begin{figure} 
	\centering
	\includegraphics[width=0.3\textwidth]{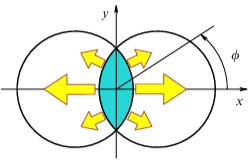}
	\caption{Collision of two heavy nuclei as seen along the collision axis. Pressure gradients along $x$-axis are shown to be greater than along $y$-axis. Source: \textcite{Ar:2014}}
	\label{fig:Fig2}
\end{figure}

\subsubsection{Relativistic Hydrodynamics}

L. Landau pioneered the use of ultrarelativistic hydrodynamics in describing multiparticle production from collisions of nuclei. The model describes QGP medium as a perfect fluid which was further simplified by Bjorken for the case of boost invariant flow. In Bjorken's simplistic model, the system undergoes one dimensional expansion along the collision axis and has boost invariance along this axis. Bjorken's model provides a further generalization of the perfect fluid case by incorporating shear/bulk viscosities and thermal conductivity for dissipative flows \autocite{Nak:2006}. In the late time regime, when the microscopic degrees of freedom have relaxed, such a hydrodynamic approximation can be readily applied to QGP dynamics. In Bjorken's model, the stress-energy-momentum tensor $T^{\mu\nu}$ takes the familiar hydrodynamic form
\begin{equation}
	T^{\mu\nu}=(\varepsilon+P)u^\mu u^\nu+Pg^{\mu\nu}+\Pi^{\mu\nu}
	\label{eq:Eq1}
\end{equation}

\noindent where $\varepsilon$ is the local energy density, $P$ is pressure and $u^\mu$ is the local 4-velocity. $\Pi^{\mu\nu}$ is the sum of dissipative contributions to the energy-momentum tensor. Up to first order in gradients it is given by
\begin{equation}
	\Pi^{\mu\nu}=-\eta\left(\Delta^{\mu\lambda}\nabla_{\lambda}u^\nu+\Delta^{\nu\lambda}\nabla_{\lambda}u^\mu-\frac{2}{3}\Delta^{\mu\nu}\nabla_{\lambda}u^\lambda\right)-\zeta\Delta^{\mu\nu}\nabla_{\lambda}u^\lambda, \qquad\qquad \Delta^{\mu\nu}=g^{\mu\nu}+u^\mu u^\nu
	\label{eq:Eq2}
\end{equation}

\noindent where $\eta$ is the shear viscosity and $\zeta$ is the bulk viscosity. $T^{\mu\nu}$ is diagonal in the local rest frame given by coordinate transformation $\{t, x_1, x_2, x_3\}\equiv\{\tau \cosh(y),\tau \sinh(y), x2, x3\}\rightarrow\{\tau,y=0, x2, x3\}$ for collision along the $x_1$ axis. Here $\tau$ is the proper time and $y$ is the rapidity. In this frame,
\begin{equation}
	T^{\mu\nu}=\begin{pmatrix} 
		\varepsilon & 0 & 0 & 0 \\
		0 & \frac{1}{\tau^2}\left(P-\frac{4\eta}{3\tau}-\frac{\zeta}{\tau}\right) & 0 & 0 \\
		0 & 0 & P+\frac{2\eta}{3\tau}-\frac{\zeta}{\tau} & 0 \\
		0 & 0 & 0 & P+\frac{2\eta}{3\tau}-\frac{\zeta}{\tau}
	\end{pmatrix}
	\label{eq:Eq3} 
\end{equation}

\noindent For a 4-dimensional Conformal Field Theory (CFT), energy-momentum conservation $\nabla_\mu T^{\mu\nu}=0$ yields
\begin{equation}
	T^{\mu\nu}=\begin{pmatrix} 
		\varepsilon & 0 & 0 & 0 \\
		0 & \frac{1}{\tau^2}\left(-\varepsilon-\tau\frac{\partial\varepsilon}{\partial\tau}\right) & 0 & 0 \\
		0 & 0 & \varepsilon+\frac{1}{2}\tau\frac{\partial\varepsilon}{\partial\tau} & 0 \\
		0 & 0 & 0 & \varepsilon+\frac{1}{2}\tau\frac{\partial\varepsilon}{\partial\tau}
	\end{pmatrix}
	\label{eq:Eq4} 
\end{equation}

From (\ref{eq:Eq3}) and (\ref{eq:Eq4}), one finds the equation of state to be $\varepsilon=3P$ with $\varepsilon\sim\tau^{-4/3}$ as $\tau\rightarrow\infty$. Although this approach is quite successful in explaining late time QGP dynamics, it cannot be used to study QGP formation from nuclei collisions as such a system is far away from equilibrium. Also, the hydrodynamic equations are highly susceptible to fluctuations of initial data making it highly non-trivial to compute transport coefficients. A radically different yet highly successful approach to understanding QGP dynamics derives from the AdS\textsubscript{5}/CFT\textsubscript{4} correspondence.  

\subsection{The AdS\textsubscript{5}/CFT\textsubscript{4} Correspondence}

AdS\textsubscript{5}/CFT\textsubscript{4} establishes a correspondence between parameters in a strongly coupled conformal field theory on the 4-d Minkowski boundary of a 5-d AdS\textsubscript{5} (anti-deSitter) space with those in a dual weakly interacting gravitational theory in its 5-d bulk \autocite{Mal:2000}. It has been highly successful in describing QGP formation, its properties in thermal equilibrium as well as its transport coefficients close to equilibrium. This approach has also been employed recently to make predictions about far from equilibrium QGP by solving the dual gravitational problem in the bulk. Using the duality, the formation of QGP and thermalization can be described as being dual to black brane formation in the AdS\textsubscript{5} space with temperature given by the Hawking temperature. A known example of such correspondence is the $\mathcal{N}= 4$ supersymmetric Yang-Mills theory, with matter fields in the adjoint representation of the SU($N_c$) gauge group, that is dual to the type IIB superstring theory in AdS\textsubscript{5}$\times$S\textsubscript{5} space \autocite{Mal:2000}. Before delving in the details, it is worthwhile mentioning the limitations of this approach.

One of the hurdles in generalizing this approach, especially to non-abelian plasma far away from equilibrium is that the dual Einstein Field Equations (EFEs) turn out to be highly coupled and do not always yield closed form analytical solutions \autocite{Sol:2014}. In a special class of problems, symmetries of the system are exploited to yield insight into the hydrodynamic behavior of QGP close to equilibrium. These are based on evaluating the gravity dual to Bjorken’s boost invariant flow in relativistic hydrodynamics described in the previous section.

A more fundamental problem with this approach is that AdS\textsubscript{5}/CFT\textsubscript{4} correspondence is exact only between a supersymmetric conformal field theory, i.e. $\mathcal{N}= 4$ Supersymmetric Yang Mills theory on the AdS\textsubscript{5} boundary and classical Supergravity in the bulk for group rank $N_c\to \infty$ and t'Hooft coupling $\lambda\to\infty$. QCD is neither super-symmetric nor conformal in general, i.e., its coupling constant changes with the energy scale in consideration. Also, quarks in QCD transform under the fundamental representation of color gauge group SU(3) whereas charges in $\mathcal{N}= 4$ SYM transform under the adjoint representation of the group SU($N_c$). However, lattice gauge theory calculations suggest that in the deconfined QGP regime, QCD is quasi conformal from the standpoint of thermodynamics. On the other hand, it has been shown how supersymmetry in $\mathcal{N}= 4$ SYM is absent for collective excitations at finite non-zero temperature using Keldysh diagram techniques \autocite{Ar:2014,Sol:2014}. Owing to these observations and experimental data from heavy ion collisions, $\mathcal{N}= 4$ SYM serves as a good model for describing QGP dynamics despite the significant differences between the two theories.

\subsubsection{AdS\textsubscript{5}/CFT\textsubscript{4} Dictionary}

The metric of AdS\textsubscript{5} space is a solution of vacuum Einstein equations (\ref{eq:Eq5}) in 5 spacetime dimensions with a constant negative curvature and negative cosmological constant \autocite{Mal:2000}
\begin{equation}
	R_{mn}-\frac{1}{2}g_{mn}R-6g_{mn}=0
	\label{eq:Eq5}
\end{equation}

\noindent where $R_{mn}$ is the 5-d Ricci tensor and $g_{mn}$ the metric ($m,n = t,x,y,z,r$). Its line element has the form:
\begin{equation}
	ds^2=\frac{r^2}{L^2}\eta_{\mu\nu}dx^\mu dx^\nu+\frac{L^2}{r^2}dr^2
	\label{eq:Eq6}
\end{equation}

\noindent where $r$ is the radial coordinate orthogonal to the 4-d spacetime directions, $L$ is the curvature radius of AdS\textsubscript{5} and $\eta_{\mu\nu}$ is the 4-d Minkowski metric ($\mu,\nu = t,x,y,z$). In the low energy limit, AdS\textsubscript{5}/CFT\textsubscript{4} correspondence relates $\mathcal{N}= 4$ SYM theory, with gauge group SU($N_c$), and Type IIB (closed) string theory in AdS\textsubscript{5}$\times$S\textsubscript{5}. The parameters characterizing the gauge theory are rank of the gauge group $N_c$ and the t’Hooft coupling $\lambda=g^2_{YM}N_c$. On the other side of the duality, parameters characterizing Type IIB string theory are the string coupling $g_s$ and the string length $l_s$. AdS\textsubscript{5}/CFT\textsubscript{4} relates these parameters as under
\begin{equation}
	g_s\sim \frac{g^2_{YM}}{4\pi} \qquad\qquad \frac{L}{l_s}\sim\lambda^{1/4} \qquad\qquad \frac{G_{10}}{L^8}\sim\frac{\pi^4}{2N^2_c}
	\label{eq:Eq7}
\end{equation}

\noindent where $G_{10}$ is the 10-d gravitational constant and $g_{YM}$ is the Yang-Mills coupling in $\mathcal{N}= 4$ SYM. 

In the limit of $\lambda\rightarrow\infty$, $N_c\rightarrow\infty$, the boundary gauge theory is strongly coupled while closed strings in dual Type IIB theory are weakly coupled (from (\ref{eq:Eq7})). In this regime, Type IIB string theory simplifies to Type IIB supergravity which on compactification over $S_5$ reproduces classical Einstein field equations coupled to matter fields. In absence of matter fields, the solution to compactified supergravity equations is simply the AdS\textsubscript{5} metric (\ref{eq:Eq6}) as expected in classical general relativity.

A $p$-form bulk field $\Phi(r,x)$ dual to an operator $O(x)$ of the boundary gauge theory deforms the gauge theory action \autocite{Sol:2014} as
\begin{equation}
	S\rightarrow S+\int{d^4 x\,\phi(x) O(x)}, \qquad\qquad \phi(x)=\lim_{r \to \infty} r^{4-p-\Delta}\, \Phi(r,x)
	\label{eq:Eq8}
\end{equation}

\noindent where $\phi(x)$ is like a source field in the gauge theory Lagrangian. $\Delta$ is the conformal dimension of its dual gauge theory operator $O(x)$ given by the larger root of
\begin{equation}
	m^2 L^2=(\Delta-p)(\Delta+p-4)
	\label{eq:Eq9}
\end{equation} 

\noindent where $m$ is the mass of the gauge theory field. Defining coordinate $z=\frac{L^2}{r}$, the near boundary expansion of the source field $\Phi(z,x)$ is
\begin{equation}
	\Phi(z,x)\equiv\Phi(z,x)_{\mu_0\dots\mu_p}\approx \alpha(x)_{\mu_0\dots\mu_p}\,z^{4-p-\Delta}+\beta(x)_{\mu_0\dots\mu_p}\,z^{\Delta-p}
	\label{eq:Eq10}	
\end{equation} 

Based on the duality prescription, it can be shown that the expectation value of operator $O(x)$ in the presence of source field $\phi(x)=\alpha(x)$ is given by (indices not shown for brevity) 
\begin{equation}
	\langle O(x) \rangle_\phi = \lim_{z \to 0} z^{4-p-\Delta} \frac{\delta S^{\mathrm{(ren)}}[\Phi_c]}{\delta \Phi_c(z,x)}=2v\beta(x), \qquad\qquad v=\sqrt{(2-p)^2+m^2 L^2}
	\label{eq:Eq11}
\end{equation}

\noindent where $S^{\mathrm{(ren)}}$ is the renormalized bulk action and $\Phi_c(z,x)$ is the solution of the classical equations of motion $\delta S[\Phi_c]=0$. To illustrate, consider a small perturbation $h_{\mu\nu}$ of the Minkowski boundary metric $\eta_{\mu\nu}$. Using (\ref{eq:Eq11}), one can compute the boundary stress-energy tensor
\begin{equation}
	\langle T_{\mu\nu}(x) \rangle_h = \lim_{z \to 0} \frac{1}{4!}\cdot\frac{1}{4\pi G_5} \frac{\partial^4 g_{\mu\nu}(z,x)}{\partial z^4} = \lim_{z \to 0} \frac{1}{4!}\cdot\frac{N^2_c}{2\pi^2 L^3} \frac{\partial^4 g_{\mu\nu}(z,x)}{\partial z^4}, \qquad\qquad g_{\mu\nu}=\eta_{\mu\nu}+h_{\mu\nu}
	\label{eq:Eq12}
\end{equation}

\subsubsection{Equilibrium QGP in AdS\textsubscript{5}/CFT\textsubscript{4}}

The geometry dual to $\mathcal{N}= 4$ SYM plasma at finite non-zero temperature is the AdS\textsubscript{5}-Schwarschild black brane \autocite{Sol:2014} with the following metric (line element)
\begin{equation}
	z^2\,ds^2 = -\left( 1-\frac{z^4}{z^4_0} \right) dt^2 + d\mathrm{\bf{x}}^2 + \frac{dz^2}{\left( 1-\frac{z^4}{z^4_0} \right)}
	\label{eq:Eq13}
\end{equation}

The corresponding Hawking temperature is $T_H = \frac{1}{\pi z_0}$. With a coordinate transformation $z \to \tilde{z}\left( 1+\tilde{z}^4/\tilde{z}^4_0 \right)^{-1/2}$, one can write (\ref{eq:Eq13}) in the Fefferman-Graham form as
\begin{equation}
	\tilde{z}^2\,ds^2 = -\frac{\left( 1-\frac{\tilde{z}^4}{\tilde{z}^4_0} \right)^2}{\left( 1+\frac{\tilde{z}^4}{\tilde{z}^4_0} \right)} dt^2 + \left( 1+\frac{\tilde{z}^4}{\tilde{z}^4_0} \right) d\mathrm{\bf{x}}^2 + d\tilde{z}^2
	\label{eq:Eq14}
\end{equation}

From (\ref{eq:Eq12}) and (\ref{eq:Eq14}), the stress-energy-momentum tensor for QGP in equilibrium corresponds to that of a perfect fluid in hydrodynamics ($\varepsilon=3P$)
\begin{equation}
	T_{\mu\nu}=\frac{N^2_c}{2\pi^2}\mathrm{\bf{diag}} \left\{ \frac{3}{\tilde{z}^4_0}, \frac{1}{\tilde{z}^4_0}, \frac{1}{\tilde{z}^4_0}, \frac{1}{\tilde{z}^4_0} \right\}
	\label{eq:Eq15} 
\end{equation}

\subsection{Quasinormal Modes of Black Branes}

Quasinormal modes (QNMs) are analogues of normal modes (characteristic oscillations) for perturbed physical systems in the presence of dissipation \autocite{Star:2009}. More formally, they are the eigenstates of linearized equations of motion for dissipative dynamical systems under perturbation. Gravitational backgrounds such as black holes/branes are intrinsically dissipative due to the presence of an event horizon. Such a system is not time-symmetric and the associated boundary value problem is not Hermitian. In general, their QNMs have complex frequencies with the imaginary part being associated with the decay timescale of the perturbation.   

In the context of gauge-gravity duality, the quasinormal spectra of dual gravitational backgrounds provide the locations of poles of the retarded correlators in the boundary gauge theory \autocite{Mal:2000,Star:2009}. They provide important information about the gauge theory's quasiparticle spectra and transport coefficients. QNMs are therefore a powerful tool in studying the near-equilibrium behavior of strongly coupled non-abelian plasmas with a dual gravitational description.

In general relativity, linearized Einstein field equations for perturbed gravitational backgrounds naturally give rise to QNMs when transformed to the Fourier space. The perturbations obey linear second-order partial differential equations (PDEs) which may be reduced to linear ordinary differential equations (ODEs) by exploiting symmetries of the gravitational background under consideration. Under appropriate boundary conditions, eigenmodes of the system of ODEs can be computed which are QNMs of the gravitational background \autocite{Star:2009}.   

\subsubsection{Computation of QNMs}

Asymptotically AdS\textsubscript{5} black brane metrics with SO(3) rotational, translational and boost symmetry have a line element of the form \autocite{Star:2009,Yaf:2015}
\begin{equation}
	ds^2_5 = \frac{r^2}{L^2}\left( -f(r)dt^2+d\textbf{x}^2 \right) + \frac{L^2}{r^2}f(r)^{-1} dr^2
	\label{eq:Eq16}
\end{equation}

\noindent where the metric function $f(r)$ encodes the bulk geometry and horizon information. For the AdS\textsubscript{5}-Schwarschild black brane $f(r)=\left(1-r^4_0/r^4\right)$, where $r_0$ is the horizon location ($f(r_0)=0$). The metric (\ref{eq:Eq16}) becomes singular at the horizon owing to the above choice of coordinates \autocite{Star:2009}. One can circumvent this issue by defining the tortoise coordinate $r^*$ related to the radial coordinate $r$ as $\frac{dr^*}{dr}=\frac{L^2}{r^{2}f(r)}$. \newline

\noindent The general prescription to compute the QNMs of (\ref{eq:Eq16}) is as follows \autocite{Star:2009,Kam:2016}

\begin{itemize}
	\item{Consider a plane wave perturbation of the metric $h_{mn}(r)\sim e^{-i\omega(t+r^*)+i\vec{k}\cdot\vec{x}}$. Using the tortoise coordinate $r^*$ ensures only infalling modes are $\mathcal{C}^{\infty}$-smooth at the horizon ($r \to r_0$)}
	\item{For an SO(3) symmetric metric, it is convenient to set $\vec{x}=\vec{z}$}
	\item{At the AdS\textsubscript{5} boundary ($r \to \infty$), gauge-gravity duality dictates that the perturbation vanish. Thus $\lim_{r \to \infty}h_{mn}(r)\sim r^{-\gamma} \,\, (\gamma > 0)$}
	\item{Substitute the perturbed metric ansatz ($g_{mn}\to g_{mn}+h_{mn}$) in Einstein field equations and linearize them w.r.t $h_{mn}(r)$}
	\item{Solve the resultant eigenvalue equations for $h_{mn}(r)$ after fixing the diffeomorphism gauge and momentum $k$. The eigenvalues obtained are the complex QNM frequencies.}
\end{itemize}

The metric perturbations are grouped into three sectors depending on the symmetry channels they transform under, viz. the scalar, vector and tensor perturbations \autocite{Star:2009}. Each sector is governed by an independent closed group of differential equations. An easy choice for fixing the diffeomorphism gauge is setting $h_{rm}=0,\,m=r,t,x,y,z$ \autocite{Kam:2016}. For non-vanishing momentum, the scalar sector comprises of perturbations that transform as a scalar under SO(2) (rotations in the $x\leftrightarrow y$ plane) - $h_{tt},\,h_{xx}+h_{yy},\,h_{tz},\,h_{zz}$. Similarly, the vectorial sector comprises of those that transform as a vector under SO(2) - $h_{tx},\,h_{ty},\,h_{zx},\,h_{zy}$ and tensorial sector comprises of those that transform as a rank-2 tensor under SO(2) - $h_{xx}-h_{yy},\,h_{xy}$.

\subsubsection{QNMs of AdS\textsubscript{5}-Schwarschild Black Branes}

In this section, we will briefly illustrate the process of computing QNMs for perturbations belonging to the tensorial sector using the example of AdS\textsubscript{5}-Schwarschild black branes. A similar process can be followed to compute the vectorial and scalar sector QNMs. As discussed in the previous section, computation of QNMs can be most easily done by working in the infalling null coordinates $(v=t+r^*,x,y,z,r)$, also known as the generalized Eddington-Finkelstein coordinates. Their advantage over Fefferman-Graham coordinates is that the QNM boundary conditions only require solutions to be regular at the horizon and vanish at the AdS\textsubscript{5} boundary \autocite{Star:2009}. In these coordinates, the AdS\textsubscript{5}-Schwarschild black brane metric takes the form
\begin{equation}
	ds^2_5 = \frac{r^2}{L^2}\left( -f(r)dv^2+d\textbf{x}^2 \right) + 2dv dr
	\label{eq:Eq17}
\end{equation} 

Without loss of generality, one can set the horizon $r_0$ as the AdS\textsubscript{5} curvature radius $L$. On adding the tensorial metric perturbation $h_{xy}\equiv \frac{r^2}{L^2}g(r)e^{-i\omega v+ikz}$ to (\ref{eq:Eq17}), and linearizing the corresponding (vacuum) Einstein field equations (\ref{eq:Eq5}), we get a QNM second order ordinary differential equation for $g(r)$
\begin{equation}
	L^2 r (L^2 k^2+3ir\omega)g(r) + (L^4-5r^4+2iL^2 r^3 \omega)g'(r) + r(L^4-r^4)g''(r) = 0
	\label{eq:Eq18} 
\end{equation} 

For every choice of momentum $k$, solutions to (\ref{eq:Eq18}) only exist for a discrete set of values of $\omega$. It is therefore an eigenvalue equation where the solutions are the eigenvectors and corresponding $\omega$'s are the eigenvalues. Solving (\ref{eq:Eq18}) using pseudospectral methods will be covered in the next few sections. For the case of vanishing momentum $k=0$ and $L=1$, the lowest few QNM frequencies ($\omega$) for the AdS\textsubscript{5}-Schwarschild black brane (\ref{eq:Eq17}) are tabulated in Table. \ref{tab:Tb1}. All QNM frequencies have a negative imaginary part which indicates oscillations damped in time.

\begin{table}[h]
	\centering
	\begin{tabular}{c c c}
		\toprule
		n & Re($\omega_n$) & Im($\omega_n$) \\
		\midrule
		1 & $\pm 3.119452$ & $-2.746676$ \\
		2 & $\pm 5.169521$ & $-4.763570$ \\
		3 & $\pm 7.187931$ & $-6.769565$ \\
		4 & $\pm 9.197199$ & $-8.772481$ \\
		5 & $\pm 11.202676$ & $-10.774162$ \\
		6 & $\pm 13.206247$ & $-12.775239$ \\
		\bottomrule
	\end{tabular}
	\caption{Lowest QNM frequencies (tensorial sector) for AdS\textsubscript{5}-Schwarschild black brane ($k=0$)}
	\label{tab:Tb1}
\end{table}

\section{Homogenous Isotropization in \( \mathcal{N} \)= 4 SYM Plasma}

It is computationally challenging to study real time dynamics of strongly coupled QGP using lattice methods. In order to study QGP formation from heavy ion collisions and its dynamics far from equilibrium, one can study the behavior of $\mathcal{N}= 4$ SYM plasma as a model of QGP for aforementioned reasons. A high degree of momentum anisotropy would seem to exist in partons formed out of such collisions, which would subsequently form the QGP state that behaves like a relativistic fluid. When this QGP cools below the crossover temperature, hadrons form that fly outwards toward the detectors. Although the near-equilibrium behavior of QGP is explained well by hydrodynamics, the same cannot be said for the rapid isotropization and mechanisms underlying QGP formation far from equilibrium. This process can be studied for the $\mathcal{N}= 4$ SYM plasma far from equilibrium using its dual gravitational description for a given set of initial/boundary conditions. The next few sections provide an example application of pseudospectral methods for solving dual Einstein field equations for the case of homogenous isotropization of far from equilibrium QGP. We use the setup and methods described in article \autocite{Yaf:2014}     

\subsection{Gravitational Description}

The most general asymptotically AdS\textsubscript{5} metric with translational and boost invariance (in Eddington-Finkelstein coordinates) has line element \autocite{Yaf:2014}
\begin{equation}
	ds^2_5 = g_{ij}(r,v) dx^i dx^j + 2 dv (dr - A(r,v)dv)
	\label{eq:Eq19}	
\end{equation}

\noindent where $i,j=x,y,z$. For a fluid expansion restricted to the $z$ direction, one can set $g_{ij}\equiv\{\Sigma^2 e^B,\Sigma^2 e^B,\Sigma^2 e^{-2B}\}$. The corresponding Einstein field equations (\ref{eq:Eq5}) are:
\begin{eqnarray}
	&\Sigma ''+\frac{1}{2}\left(B'\right)^2 \Sigma = 0
	\label{eq:Eq20}\\
	&A'' = 6\left(\frac{\Sigma'}{\Sigma^2}\right)d_+ \Sigma - \frac{3}{2}B' d_+ B - 2
	\label{eq:Eq21}\\
	&d_+ \Sigma' + 2\left(\frac{\Sigma'}{\Sigma}\right)d_+ \Sigma = 2\Sigma
	\label{eq:Eq22}\\
	&\left(d_+ B\right)' + \frac{3}{2}\left(\frac{\Sigma'}{\Sigma}\right)d_+ B = -\frac{3}{2}B'\frac{\left(d_+ \Sigma\right)}{\Sigma}
	\label{eq:Eq23} 
\end{eqnarray}

\noindent where $'\equiv \partial_r$ and $d_+\equiv \partial_v + A(r_*,v_*)\partial_r$ is the directional derivative along the outgoing null geodesic passing through $(r_*,v_*)$. Before solving the system of equations ($\ref{eq:Eq20}\to\ref{eq:Eq23}$), one needs to fix the residual gauge freedom. One computationally convenient way of doing this, for reasons mentioned in the next section, is demanding that the location of apparent horizon (outermost trapped null surface) occurs at a fixed value of radial coordinate $r=r_h$ and that it does not change with time \autocite{Yaf:2014}. This requirement leads to two additional constraints for metric functions
\begin{eqnarray}
	&d_+ \Sigma\big|_{r_h} = 0
	\label{eq:Eq24}\\
	&A(r_h,v) = -\frac{1}{4}\left(d_+ B\right)^2\big|_{r_h}
	\label{eq:Eq25} 
\end{eqnarray}

The metric form (\ref{eq:Eq19}) is manifestly invariant under a radial shift $r\to r+\lambda(v)$. In effect, (\ref{eq:Eq24}) fixes the initial value of radial shift $\lambda(v_0)$ at time $v_0$ and (\ref{eq:Eq25}) determines $\partial_v\lambda$ to keep the horizon stationary. The asymptotic near boundary expansions ($r\to\infty$) for the metric functions are found to be
\begin{eqnarray}
	&\Sigma \sim r + \lambda(v) + O(r^{-7})
	\label{eq:Eq26}\\
	&A \sim \frac{1}{2}\left(r+\lambda(v)\right)^2 - \partial_v\lambda(v) + a^{(4)}(v)r^{-2} + O(r^{-3})
	\label{eq:Eq27}\\
	&d_+ \Sigma \sim \frac{1}{2}\left(r+\lambda(v)\right)^2 + a^{(4)}(v)r^{-2} + O(r^{-3})
	\label{eq:Eq28}\\
	&B \sim b^{(4)}(v)r^{-4} + O(r^{-5})
	\label{eq:Eq29}\\
	&d_+ B \sim -2b^{(4)}(v)r^{-3} + O(r^{-4})
	\label{eq:Eq30}
\end{eqnarray}

\noindent where $a^{(4)}(v)$ and $b^{(4)}(v)$ are related to the boundary stress-energy tensor based on the gauge-gravity duality as (parantheses omitted for brevity)
\begin{equation}
	T^{vv} = -\frac{3N_c^2}{4\pi^2}a^{(4)}, \qquad\qquad T^{xx} = T^{yy} = \frac{N_c^2}{2\pi^2}\left(b^{(4)}-\frac{1}{2}a^{(4)}\right), \qquad\qquad T^{zz} = \frac{N_c^2}{2\pi^2}\left(-2b^{(4)}-\frac{1}{2}a^{(4)}\right)
	\label{eq:Eq31}
\end{equation}

It can be seen that (\ref{eq:Eq26}), (\ref{eq:Eq27}) and (\ref{eq:Eq28}) diverge as $r\to\infty$. It is computationally convenient to use the inverse radial coordinate $u=1/r$ and define subtracted and rescaled functions (i.e. functions with leading order $r$ divergences removed and subsequently rescaled) such that these quantities approach a constant or vanish linearly as $u\to 0$ \autocite{Yaf:2014}. Following is an example of such redefinition from article \autocite{Yaf:2014}
\begin{equation}
	\sigma = \Sigma - 1/u, \qquad\qquad \dot{\sigma} = u^{-1}\left(d_+ \Sigma - \frac{1}{2}\Sigma^2\right), \qquad\qquad a = A - \frac{1}{2}\Sigma^2
	\label{eq:Eq32}
\end{equation} 
\begin{equation}
	b = u^{-3}B, \qquad\qquad \dot{b}=u^{-3}d_+ B
	\label{eq:Eq33}
\end{equation}

\noindent where resulting $u\to 0$  boundary conditions to be imposed for these redefined fields are
\begin{equation}
	\sigma\to \lambda, \qquad \dot{\sigma}\to ua^{(4)}, \qquad b\to 0, \qquad \dot{b}\to -2b^{(4)}
	\label{eq:Eq34}
\end{equation}

\subsection{Numerics}

The system of linear ordinary differential equations (ODEs) ($\ref{eq:Eq20}\to\ref{eq:Eq23}$) has a very convenient nested structure. Given the anisotropy function $B(v_0,r)$, boundary energy density $T^{vv}(v_0)$ and radial shift $\lambda(v_0)$ at some initial time $v_0$, one can solve the equations one after the other in order $(\ref{eq:Eq20})\to(\ref{eq:Eq22})\to(\ref{eq:Eq23})\to(\ref{eq:Eq21})$ such that constraints (\ref{eq:Eq24}) and (\ref{eq:Eq25}) are satisfied. Using the definition of $d_+ B$, the function $B$ can be evolved in time and the aforementioned process is repeated to get the time evolution of the other metric functions. As discussed in the previous section, we choose to work with the redefined fields ($\ref{eq:Eq26}\to\ref{eq:Eq30}$) for computational reasons, with boundary conditions (\ref{eq:Eq34}). 

The initial value of radial shift $\lambda(v_0)$ is not known apriori and must be computed using a root finding technique such as Newton-Raphson method to solve constraint (\ref{eq:Eq24}) in conjunction with solving (\ref{eq:Eq20}) and (\ref{eq:Eq21}). The next section describes how pseudospectral methods (esp. those employing Chebyshev basis functions) can be used to solve linear boundary value problems (BVPs).

\subsubsection{Pseudospectral ODE Solvers}

Traditionally a system of ODEs is solved numerically using some variant of Finite Element Method (FEM). It involves discretization of the domain space ($r$ in this case) into a grid. The functions are represented as piecewise polynomial basis functions defined on the grid. This turns the BVPs into a system of linear equations, one per each domain element/piece per function. The derivatives of functions are approximated by the local derivatives of the basis functions. These methods work quite well when there are no singular points in the computational domain (including the domain boundary).

For the system described in the previous section, we can see that $u\to 0$ and $u\to 1$ are singular points which are not well tolerated by FEM. One can circumvent this issue by using spectral methods for solving ODEs as one can directly apply spectral methods to BVPs with regular singular points as long as the solution of interest is regular at the singular point \autocite{Boyd,Yaf:2014}. Also, they show superior convergence as compared to FEM and their accuracy improves exponentially with the number of basis functions. Unlike FEM, spectral methods use a truncated linear combination of long range basis functions which are defined over the entire computational domain of interest. As with FEM, spectral methods also transform a BVP into a linear algebraic problem. For periodic functions defined on an interval, the basis of complex exponentials (Fourier basis) is a natural choice and the expansion is the truncated Fourier series \autocite{Yaf:2014}. For aperiodic functions, the convenient basis is the Chebyshev polynomials $T_n(z)\equiv \cos(n \cos^{-1}(z))$. For a computational domain $0<u<1$, the (truncated) expansion of a function $f(u)$ is given by
\begin{equation}
	f(u) \approx \sum^N_{n=0} \alpha_n T_n(2u-1)
	\label{eq:Eq35}
\end{equation}    

Solving a BVP for $f(u)$ reduces to solving for the coefficients $\alpha_n$ after substituting (\ref{eq:Eq35}) into the BVP. In pseudospectral or collocation approaches, one makes such a substitution and demands that the residual of the BVP vanishes exactly at a predetermined special set of points called the collocation grid points \autocite{Boyd,Yaf:2014}. The number of such points equals the expansion order $N$. In this approach, the knowledge of coefficients $\alpha_n$ is equivalent to the value of function $f(u)$ at collocation grid points points $u_n$. For the Chebyshev basis, appropriate collocation points in the domain interval $[0,1]$ are
\begin{equation}
	u_n = \frac{1}{2}\left(1 - \cos{\frac{n\pi}{N}}\right)
	\label{eq:Eq36}
\end{equation} 

\subsubsection{Time Evolution}

For each timestep, once the system of (radial) ODEs is solved using a pseudospectral solver, the anisotropy function $B$ can be evolved in time by solving the Initial Value Problem (IVP) $\partial_v B = d_+ B - AB'$ \autocite{Yaf:2014}. This can be solved using an explicit fourth order Runge-Kutta (RK4) method which has a convergence error of $O(\Delta v)^5$ in the timestep $\Delta v$. In RK4, for an IVP of the form
\begin{equation}
	\frac{\partial f}{\partial v} = g(v, f), \qquad\qquad f(v_0) = f_0
	\label{eq:Eq37}
\end{equation}

\noindent and a chosen timestep size of $\Delta v\equiv h>0$, the function $f(v+h)$ can be approximated by
\begin{equation}
	f(v+h) \approx f(v) + \frac{h}{6}\left(k_1+2k_2+2k_3+k_4\right)
	\label{eq:Eq38}
\end{equation} 

\noindent where,
\begin{eqnarray}
	&k_1 = g(v, f)
	\label{eq:Eq39}\\
	&k_2 = g\left(v+\frac{h}{2}, f+h\frac{k_1}{2}\right)
	\label{eq:Eq40}\\
	&k_3 = g\left(v+\frac{h}{2}, f+h\frac{k_2}{2}\right)
	\label{eq:Eq41}\\
	&k_4 = g(v+h, f+hk_3)
	\label{eq:Eq42}
\end{eqnarray}

\subsection{Results}

The following was chosen as the initial data and parameters for the example simulation
\begin{equation}
	\nonumber b(v_0, u) = 5ue^{-25\left(u-\frac{1}{4}\right)^2}, \qquad\qquad a^{(4)} = -\frac{1}{2}, \qquad\qquad r_h = 1, \qquad\qquad N = 51, \qquad\qquad h = 10^{-2}
\end{equation}

Plot (Fig. 3) shows the time evolution of the boundary pressure anisotropy $\delta P\equiv T_{zz}-\frac{1}{2}\left(T_{xx}+T_{yy}\right)$ in units of $\frac{N^2_c}{2\pi^2}$ for $v_0\leq v<v_0+4$. It's value can be computed from (\ref{eq:Eq31}) to be $\delta P = -3b^{(4)}\frac{N^2_c}{2\pi^2}$. As seen from the plot, the pressure anisotropy rapidly decays with exponentially damped oscillations till equilibrium is reached, i.e. till $T_{xx}=T_{yy}=T_{zz}=\frac{N^2_c}{8\pi^2}$ where the plasma is homogenous and isotropic. Plot (Fig. 4) shows the computed total anisotropy function in the AdS\textsubscript{5} bulk - $B(u,v)$ for $0\leq u\leq 1$ and $v_0\leq v<v_0+4$. It also decays exponentially with time for all $u$ till the geometry becomes that of an AdS\textsubscript{5}-Schwarschild black brane with horizon fixed at $r_h=1$ and Hawking temperature $T_H=\frac{1}{\pi}$.

\begin{figure} 
	\centering
	\includegraphics[width=0.7\textwidth]{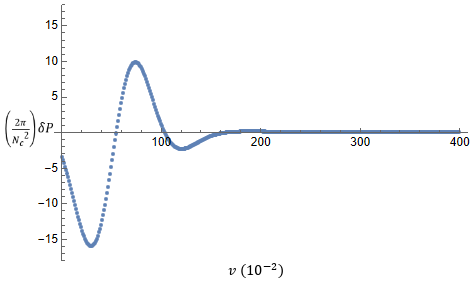}
	\caption{Normalized pressure anisotropy $\frac{2\pi^2}{N^2_c}\cdot\delta P$ shown as a function of (infalling-coordinate) time $v$.}
	\label{fig:Fig3}
\end{figure}

\begin{figure} 
	\centering
	\includegraphics[width=0.7\textwidth]{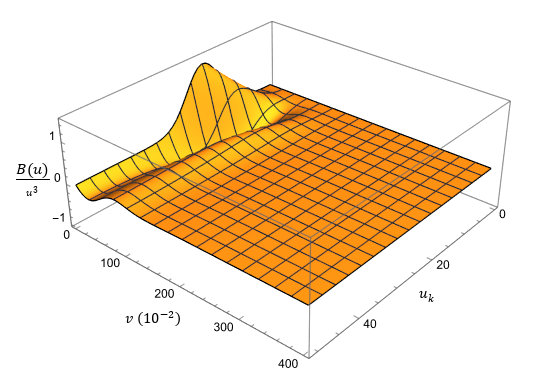}
	\caption{Time evolution of the anisotropy function $B(u,v)$ in units of $u^3$.}
	\label{fig:Fig4}
\end{figure}

Having seen an illustrative example of how to use pseudospectral methods for solving the Einstein field equations, the next two sections describe how to use this approach for computing the QNMs of a) Reisnner-Nordstrom (electrically charged) and b) Magnetically charged AdS\textsubscript{5} black branes.

\section{Reissner-Nordstrom Black Branes}

In this section, we consider the gravitational dual of an electrically charged $\mathcal{N}= 4$ SYM plasma with uniform charge density $C$ \autocite{Kam:2016,Yaf:2015}. The dual theory has a corresponding non-zero gauge field $A_m\equiv \{A_t,0,0,0\}$, which from (\ref{eq:Eq10}) is
\begin{equation}
	A_t = \mu - \frac{C}{r^2}
	\label{eq:Eq43}
\end{equation}

\noindent in units $\{c,G_5,k_e,L=1\}$ ($k_e$ is the Cuolomb constant). The corresponding (bulk) electromagnetic stress-energy tensor is 
\begin{eqnarray}
	\nonumber &{T^m}_n = \frac{1}{4\pi}\left(F^{m\alpha}F_{n\alpha} - \frac{1}{4}{\delta^m}_n F_{\alpha\beta}F^{\alpha\beta}\right)\\
	&{T^v}_v={T^r}_r=-\frac{1}{4\pi}\frac{2C^2}{r^6}, \qquad\qquad {T^i}_j=\delta_{i,j}\,\frac{1}{4\pi}\frac{2C^2}{r^6} \qquad (i,j=x,y,z)
	\label{eq:Eq44} 
\end{eqnarray}   

\noindent The Einstein-Maxwell field equations are
\begin{equation}
	R_{mn}-\frac{1}{2}g_{mn}R-6g_{mn} = 8\pi T_{mn}
	\label{eq:Eq45}
\end{equation}

\noindent which yield the following asymptotically-AdS\textsubscript{5} Reisnner-Nordstrom black brane metric (in Eddington-Finkelstein coordinates)
\begin{equation}
	ds^2_5 = r^2\left(-f(r)dv^2+d\mathrm{\bf{x}}^2\right)+2dvdr
	\label{eq:Eq46}
\end{equation}

\noindent where the metric function $f(r)$ encodes the bulk geometry and horizon information (\ref{eq:Eq47}). Two radially separated horizons exist with locations given by the positive roots of $f(r)$. The larger root is the radial location of event horizon and the smaller root is that of an inner Cauchy horizon. The black brane has an extremal value of charge density $\rho_\mathrm{ex}=\sqrt{2}\left(\frac{m}{3}\right)^{\frac{3}{4}}$ where the two roots coincide and the Hawking temperature vanishes $T_H=0$ (from (\ref{eq:Eq48})) 
\begin{equation}
	f(r) = 1 - \frac{m}{r^4} + \frac{\rho^2}{r^6}, \qquad\qquad \rho=\frac{2}{\sqrt{3}}C
	\label{eq:Eq47}
\end{equation}

By requiring regularity of the Euclidean manifold at the event horizon ($r=r_+$), the Hawking temperature $T_H$ and chemical potential $\mu$ can be computed as \autocite{Mal:2000}
\begin{equation}
	T = r^2_+ \frac{\left|f'(r_+)\right|}{4\pi}, \qquad\qquad \mu = \frac{C}{r^2_+}
	\label{eq:Eq48}
\end{equation}

\subsection{Numerics}

The horizon location $r_+$ demarcates a boundary for the computational domain as the metric (\ref{eq:Eq46}) becomes singular at that point \autocite{Yaf:2015}. In order to use the pseudospectral method for computing the quasinormal mode frequencies for this metric, it is convenient to transform the radial coordinate as $r\to \frac{r_+}{u}$. This brings the computational domain from $r\in (r_+,\infty)$ to $u\in (0, 1)$ which is the fundamental domain for Chebyshev expansions of the metric functions. The metric itself transforms as
\begin{equation}
	ds^2_5 = \frac{r^2_+}{u^2}\left(-f(u)dv^2+d\mathrm{\bf{x}}^2\right)-2\frac{r_+}{u^2}dvdu
	\label{eq:Eq49}
\end{equation}

\noindent with the temperature
\begin{equation}
	T_H = r_+ \frac{\left|f'(u=1)\right|}{4\pi} =  r_+\frac{\left(2-\rho^2r^{-6}_+\right)}{2\pi}
	\label{eq:Eq50}
\end{equation}

\subsubsection{QNMs of Reisnner-Nordstrom Black Branes}

Here we focus exclusively on gravitational fluctuations for vanishing spatial momenta ($k=0$) as they are radially infalling and correspond to pressure anisotropy in the boundary stress-energy tensor. For this case, only the rank-2 tensor metric perturbations have non-trivial QNMs, while scalar and vector perturbations only yield trivial solutions completely determined by boundary data \autocite{Kam:2016}. With $k=0$ and fixing the diffeomorphism gauge $h_{rm}=0$, independent tensor fluctuations under the full SO(3) rotation group are - $h_{xy},\,h_{yz},\,h_{zx},\,h_{xx}-h_{yy},\,h_{yy}-h_{zz}$. All of them satisfy the same equation of motion and are decoupled from the rest. 

Consider a plane wave ansatz for the shear perturbation $h_{xy}\equiv e^{-i\omega v}g(u)u^2$. Using the perturbed metric $g_{mn}=\eta_{mn}+h_{mn}$ in Einstein-Maxwell equations (\ref{eq:Eq45}), we get the equation of motion
\begin{equation}
	g''(u)+g'(u)\frac{\left(uf'(u)+5f(u)+2i\tilde{\omega}u\right)}{uf(u)}+g(u)\frac{\left(4f'(u)+5i\tilde{\omega}\right)}{uf(u)} = 0
	\label{eq:Eq51}
\end{equation}  

\noindent where $\tilde{\omega}\equiv \omega/r_+$ is the rescaled angular frequency, which removes explicit dependence of (\ref{eq:Eq51}) on the horizon location $r_+$. As can be readily checked (\ref{eq:Eq51}) takes the form of an eigenvalue equation
\begin{equation}
	\Phi\,g(u) = i\tilde{\omega}\,g(u), \qquad\qquad \Phi\equiv \Phi(\partial_u, f(u), u)
	\label{eq:Eq52}
\end{equation}

One can now use the psuedospectral approach outlined in a previous section and approximate $g(u)$ by a truncated expansion in the Chebyshev polynomial basis over $u$ at the collocation grid points (\ref{eq:Eq35}). This transforms (\ref{eq:Eq52}) into a linear algebraic eigenvalue problem with resulting eigenvalues being the QNM frequencies. 

\subsection{Results and Analysis}

We found a grid size of 50 grid points adequate to get stable QNM frequencies with rapid convergence using pseudospectral approach. Table. \ref{tab:Tb2} lists the lowest four QNM frequencies ($\tilde{\omega}$), in order of magnitude of their imaginary part, for charge density varying from 0 to 0.9$\rho_{ex}$.
\begin{table}[h]
	\centering
	\begin{tabular}{c|c|c|c|c}
		\toprule
		$\rho/\rho_\mathrm{ex}$ & n=1 & n=2 & n=3 & n=4\\ \hline
		0.0 & $\pm 3.119452-2.746676i$ & $\pm 5.169521-4.763570i$ & $\pm 7.187931-6.769565i$ & $\pm 9.179199-8.772481i$\\
		0.1 & $\pm 3.116518-2.752218i$ & $\pm 5.163269-4.774448i$ & $\pm 7.178353-6.786078i$ & $\pm 9.184267-8.794832i$\\
		0.2 & $\pm 3.107594-2.769530i$ & $\pm 5.144302-4.808750i$ & $\pm 7.149378-6.838513i$ & $\pm 9.145265-8.866208i$\\
		0.3 & $\pm 3.092364-2.800902i$ & $\pm 5.112266-4.872204i$ & $\pm 7.101081-6.937009i$ & $\pm 9.081220-9.002002i$\\
		0.4 & $\pm 3.070567-2.851114i$ & $\pm 5.068295-4.977288i$ & $\pm 7.038750-7.104216i$ & $\pm 9.005274-9.236813i$\\
		0.5 & $\pm 3.043124-2.929579i$ & $\pm 5.022870-5.148924i$ & $\pm 6.997738-7.380419i$ & $\pm 8.993865-9.613485i$\\
		0.6 & $\pm 3.017494-3.054959i$ & $\pm 5.029894-5.417824i$ & $\pm 7.089142-7.741854i$ & $\pm 9.145420-10.01256i$\\
		0.7 & $\pm 3.033036-3.254092i$ & $\pm 5.170330-5.703165i$ & $\pm 7.255584-8.055419i$ & $\pm 9.307353-10.40596i$\\
		0.8 & $\pm 3.164806-3.470840i$ & $\pm 5.326875-5.968086i$ & $\pm 7.438105-8.452644i$ & $\pm 9.540560-10.92871i$\\
		0.9 & $\pm 3.322345-3.683569i$ & $\pm 5.557329-6.366283i$ & $\pm 7.757594-9.017011i$ & $\pm 9.946420-11.65867i$\\ 
		\bottomrule
	\end{tabular}
	\caption{Lowest four QNM frequencies (tensorial sector) for $k=0$ and $\rho$ varying from 0 to 0.9$\rho_\mathrm{ex}$}
	\label{tab:Tb2}
\end{table}

From Table. \ref{tab:Tb2}, it can be seen that up until $\rho\sim 0.5\rho_\mathrm{ex}$, the real part of each QNM mode's frequency decreases with increasing $\rho$ and starts increasing somewhere between $\rho$=0.5$\rho_\mathrm{ex}$ to 0.6$\rho_\mathrm{ex}$. The imaginary parts (damping coefficients), however, increase monotonically with $\rho$ for each QNM mode's frequency. Also, the computation is numerically stable up until $\rho=\rho_\mathrm{ex}$. 

\subsubsection{QNMs for Near Extremal Reisnner-Nordstrom Black Branes}

Plot (Fig. 5) shows behavior of imaginary part of lowest QNM frequency as the extremal value of charge density $\rho_\mathrm{ex}$ is approached $\left(\rho/\rho_\mathrm{ex}=0.9\right.$ to $\left. (1-10^{-11})\right)$. It is a log-log plot with $\log(\left|\mathrm{Im}[\tilde{\omega}-\tilde{\omega}_\mathrm{ex}]\right|)$ on the \textit{y}-axis and $\log(\pi T_H/\rho^{1/3})$ on the \textit{x}-axis represented by the blue dots. The red line shows a linear fit $y=x+c_I$. Similar result is obtained for the real part as shown in plot (Fig. 6), with the linear fit $y=x+c_R$ ($c_I$ and $c_R$ are constants) and this pattern is observed for all modes. 

\begin{figure} 
	\centering
	\includegraphics[width=0.6\textwidth]{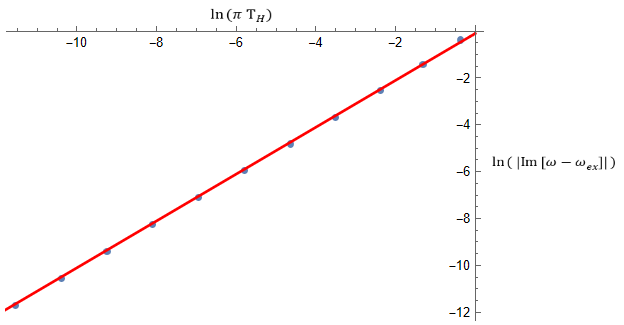}
	\caption{Plot of $\log(\left|\mathrm{Im}[\tilde{\omega}-\tilde{\omega}_\mathrm{ex}]\right|)$ vs. $\log(\pi T_H/\rho^{1/3})$ for $\rho/\rho_\mathrm{ex}=0.9$ to $(1-10^{-11})$. The red line shows a linear fit $y=x+c_I$}
	\label{fig:Fig5}
\end{figure}

\begin{figure} 
	\centering
	\includegraphics[width=0.6\textwidth]{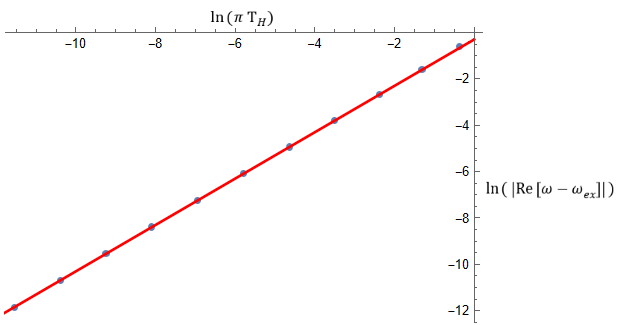}
	\caption{Plot of $\log(\left|\mathrm{Re}[\tilde{\omega}-\tilde{\omega}_\mathrm{ex}]\right|)$ vs. $\log(\pi T_H/\rho^{1/3})$ for $\rho/\rho_\mathrm{ex}=0.9$ to $(1-10^{-11})$. The red line shows a linear fit $y=x+c_R$}
	\label{fig:Fig6}
\end{figure}

We see that close to extremality, there is a linear relation between $\tilde{\omega}^{(n)}$ and $\pi T_H/\rho^{1/3}$ ($n$ is the mode \#, $\alpha_n$ are mode specific constants). From (\ref{eq:Eq53}), decay times for tensorial perturbations of near-extremal Reisnner-Nordstrom black branes are proportional only to their Hawking temperature (in units of charge density) $T_H/\rho^{1/3}$. Table. \ref{tab:Tb5} lists the values of $\alpha_n$ for the lowest ten QNM frequencies.
\begin{equation}
	\tilde{\omega}^{(n)} \approx \tilde{\omega}^{(n)}_\mathrm{ex} + \alpha_n\cdot \frac{\pi T_H}{\rho^{1/3}}, \qquad\qquad \left(\frac{T_H}{\rho^{1/3}} \ll 1\right)
	\label{eq:Eq53}
\end{equation}

\begin{table}[h]
	\centering
	\begin{tabular}{c|c}
		\toprule
		n & $\alpha_n$\\ \hline
		1 & $\mp 0.720435+0.814211i$\\
		2 & $\mp 1.203368+1.407563i$\\
		3 & $\mp 1.678953+1.993761i$\\
		4 & $\mp 2.152130+2.577889i$\\
		5 & $\mp 2.624237+3.161153i$\\
		6 & $\mp 3.095778+3.743983i$\\
		7 & $\mp 3.566986+4.326567i$\\
		8 & $\mp 4.037981+4.908998i$\\
		9 & $\mp 4.508832+5.491329i$\\
		10 & $\mp 4.979582+6.073592i$\\
		\bottomrule
	\end{tabular}
	\caption{$\alpha_n$ for lowest ten tensorial QNM frequencies ($k=0$) of Reissner-Nordstrom black branes}
	\label{tab:Tb5}
\end{table}

\section{Magnetic Black Branes}
In this section, we consider the gravitational dual of a magnetically charged $\mathcal{N}= 4$ SYM plasma with uniform magnetic field $b$ along the $\vec{z}$ direction \autocite{Yaf:2015,Kam:2016}. The dual theory has a corresponding bulk electromagnetic tensor $F_{mn}$ with non-zero components
\begin{equation}
	F_{xy} = -F_{yx} = b
	\label{eq:Eq54}
\end{equation}

Unlike the Reisnner-Nordstrom case, the metric solution to Einstein-Maxwell equations (\ref{eq:Eq45}) for this case is not analytically known. We consider an asymptotically AdS\textsubscript{5} metric ansatz (in Eddington-Finkelstein coordinates) with SO(2) symmetry in the $x\leftrightarrow y$ plane and translational/boost symmetries
\begin{equation}
	ds^2_5 = -U(r)dv^2 + e^{2V(r)}(dx^2+dy^2) + e^{2W(r)}dz^2 + 2dvdr
	\label{eq:Eq55}
\end{equation} 

\noindent The bulk electromagnetic stress-energy tensor ${T^\mu}_\nu$ as given by its non-zero components is
\begin{equation}
	{T^v}_v={T^r}_r=-b^2e^{-4V(r)}, \qquad\qquad {T^i}_i = b^2e^{-4V(r)}, \qquad (i,j=x,y,z)
	\label{eq:Eq56}
\end{equation}

The Hawking temperature $T_H$ is given by (\ref{eq:Eq48}). With a transformation of the radial coordinate $r\to 1/u$, the Einstein-Maxwell field equations (\ref{eq:Eq45}) form the system
\begin{eqnarray}
	&3U''(u)u^4+3U'(u)u^3\left(2+2uV'(u)+uW'(u)\right)-4b^2e^{-4V(u)}-24 = 0
	\label{eq:Eq57}\\
	&\left(2V''(u)+W''(u)\right)u+2\left(V'(u)^2 u+2V'(u)\right)+\left(W'(u)^2 u+2W'(u)\right) = 0
	\label{eq:Eq58}\\
	\nonumber &U'(u)\left(W'(u)-V'(u)\right)u^4+U(u)\left(W''(u)u-V''(u)u+2W'(u)-2V'(u)+W'(u)^2 u\right.\\
	&\left.-2V'(u)^2 u-2V'(u)W'(u)u\right)u^3-2b^2e^{-4V(u)} = 0
	\label{eq:Eq59}
\end{eqnarray}

We choose the horizon location $r_+=1$ by fixing the gauge choice for radial shift $r\to r+\lambda$ such that $U(u=1)=0$. Following the prescription in \autocite{Yaf:2015}, for a chosen renormalization point $\mu\sim O(\left|b\right|^{1/2})$ and a (renormalized) boundary energy density of $\varepsilon_b=3/4$, the near boundary expansions of the metric functions allow the following redefinitions. The redefined fields $\{\alpha(u),\beta(u),\sigma(u)\}$ are analytic over the computational interval $u\in(0,1)$.
\begin{eqnarray}
	&U(u)=\alpha(u)u^3-\left(1-\frac{2}{3}b^2 \log(u)-\frac{1}{3}b^2\log(b)\right)u^2+\left(\frac{1}{u}+\sigma(u)\right)^2
	\label{eq:Eq60}\\
	&V(u)=\beta(u)+\log\left(\frac{1}{u}+\sigma(u)\right)
	\label{eq:Eq61}\\
	&W(u)=-2\beta(u)+\log\left(\frac{1}{u}+\sigma(u)\right)
	\label{eq:Eq62}
\end{eqnarray}

\subsection{Numerics}
We first need to numerically compute the metric functions using the pseudospectral approach. The system of equations $\ref{eq:Eq57}\to \ref{eq:Eq59}$ are to be solved for the redefined fields $\{\alpha(u),\beta(u),\sigma(u)\}$ with AdS\textsubscript{5} boundary conditions $V(u=0)=W(u=0)=0$ and horizon fixing condition $U(u=1)=0$.

To do this, the redefined fields are approximated by a truncated expansion in Chebyshev polynomials over $u$ at the collocation grid points. As this system of equations is coupled and highly non-linear, one needs to employ a multivariate root finding algorithm (e.g. Newton's method) over the values of the redefined field at each collocation point. The newton step can be expressed in block matrix form as
\begin{equation}
	\begin{pmatrix}
		\bf{A}_{n+1}\\
		\bf{B}_{n+1}\\
		\bf{C}_{n+1}
	\end{pmatrix} =
	\begin{pmatrix}
		\bf{A}_{n}\\
		\bf{B}_{n}\\
		\bf{C}_{n}
	\end{pmatrix} -
	\begin{pmatrix}
		\bf{J}_{1A} & \bf{J}_{1B} & \bf{J}_{1C}\\
		\bf{J}_{2A} & \bf{J}_{2B} & \bf{J}_{2C}\\
		\bf{J}_{3A} & \bf{J}_{3B} & \bf{J}_{3C}
	\end{pmatrix}^{-1} \cdot
	\begin{pmatrix}
		\bf{f_1}\\
		\bf{f_2}\\
		\bf{f_3}
	\end{pmatrix}
	\label{eq:Eq63}
\end{equation} 

\noindent where $\mathrm{A}\equiv\left\{\alpha(u_1),...\alpha(u_k)\right\}$, $\mathrm{B}\equiv\left\{\beta(u_1),...\beta(u_k)\right\}$, $\mathrm{C}\equiv\left\{\sigma(u_1),...\sigma(u_k)\right\}$ and $\mathrm{J}_{ij}=\mathrm{\nabla_j}\mathrm{f_i}\big|_{\vec{u}}$ ($k$ is the grid size). $\mathrm{f_i}$ represents the computed values for ($\ref{eq:Eq57}\to\ref{eq:Eq59}$) over $u$ and boundary conditions at time step $n$. One starts with $b=0$ where the solution is AdS\textsubscript{5}-Schwarschild black brane metric (\ref{eq:Eq17}) and gradually increases $b$ so that Newton's method can converge faster \autocite{Yaf:2015}. This process is repeated till the desired magnetic field strength is reached.

\subsubsection{QNMs of Magnetic Black Branes}

As with the Reisnner-Nordstrom case, we focus exclusively on gravitational fluctuations for vanishing spatial momenta ($k=0$). The $z$ direction of the magnetic field breaks SO(3) symmetry down to SO(2) rotational symmetry in the $x\leftrightarrow y$ plane. All the three type of perturbations viz., the scalar, vector and tensor channels have non-trivial QNMs in this case. However, we focus on the scalar and tensor perturbations in this article. Because of the SO(3) to SO(2) symmetry breaking, the perturbations corresponding to boundary pressure anisotropy belong to the scalar sector for magnetic black branes \autocite{Kam:2016}. All perturbations in a sector are decoupled from other sectors and satisfy a closed set of equations of motion. 

Let us first consider the tensorial sector. We take a plane wave ansatz for shear perturbation $h_{xy}\equiv e^{-i\omega v}g(u)u^2$. Using the perturbed metric $g_{mn}=\eta_{mn}+h_{mn}$ in Einstein-Maxwell equations (\ref{eq:Eq45}), we get the equation of motion
\begin{eqnarray}
	&g''(u) + \Phi_1(u)g'(u)+\Phi_2(u)g(u) = 0\\
	\nonumber &\Phi_1(u) = \frac{6U(u)u+2i\omega+U'(u)u^2}{U(u)u^2}-2V'(u)+W'(u)\\
	\nonumber &\Phi_2(u) = \frac{2U'(u)\left(1-V'(u)u\right)u^2+i\omega\left(4-2V'(u)u+W'(u)u\right)+2U(u)\left(3+W'(u)u-V'(u)(4+W'(u)u)u+V''(u)u^2\right)u}{U(u)u^3}
	\label{eq:Eq64}
\end{eqnarray}

\noindent which is an eigenvalue equation to be solved once the metric functions are computed for a given value of $b$. The resulting eigenvalues are the QNM frequencies in the tensorial sector. 
	
Similarly consider the case of scalar perturbations. Three of the scalar perturbations $h_{vv},h_{xx}+h_{yy},h_{zz}$ are coupled in the equations of motion. We use a plane wave ansatz for each of the three scalar perturbations
\begin{equation}
	h_{vv}=e^{-i\omega v}g_1(u)u^2, \,\,\, h_{xx}=h_{yy}=e^{-i\omega v+2V(u)}\left(g_2(u)-g_3(u)\right)u^2, \,\,\, h_{zz}=-2e^{-i\omega v+2W(u)}\left(g_2(u)+g_3(u)\right)
\end{equation}

The metric ansatz perturbed with these scalar perturbations is substituted in the Einstein-Maxwell equations (\ref{eq:Eq45}) which produce three coupled equations of motion (omitted for brevity). These are eigenvalue equations and the computed eigenvalues are the QNM frequencies for the scalar sector. 

\subsection{Results and Analysis}

We used a (radial) grid of 60 points for stable convergence. Table. \ref{tab:Tb3} lists the lowest two QNM frequencies ($\omega/\pi T_H$) for the tensorial perturbations, in order of decreasing imaginary part. The (dimensionless) field strength $b/T^2_H$ varies from 0 to $\sim 34$ and boundary energy density (renormalized) is set to $\varepsilon_b=3/4$. Table. \ref{tab:Tb4} lists the lowest two QNM frequencies ($\omega/\pi T_H$) for scalar perturbations of magnetic black brane ($\varepsilon_b=3/4$).
\begin{table}[h]
	\centering
	\begin{tabular}{c|c|c}
		\toprule
		$b/T^2_H$ & n=1 & n=2\\ \hline
		0 & $\pm 3.119452-2.746676i$ & $\pm 5.169521-4.763570i$\\
		2.48 & $\pm 3.124071-2.795568i$ & $\pm 5.173049-4.843428i$\\
		5.17 & $\pm 3.132475-2.937596i$ & $\pm 5.173807-5.072583i$\\
		8.44 & $\pm 3.129093-3.178006i$ & $\pm 5.141546-5.447566i$\\
		12.95 & $\pm 3.075832-3.548793i$ & $\pm 5.006620-5.971654i$\\
		20.17 & $\pm 2.841178-4.114296i$ & $\pm 4.670453-6.541204i$\\
		34.84	& $\pm 2.021158-4.684788i$ & $\pm 4.547208-7.121793i$\\
		\bottomrule
	\end{tabular}
	\caption{Lowest two QNM frequencies (tensorial sector) ($k=0$) for magnetic black brane}
	\label{tab:Tb3}
\end{table}

\begin{table}[h]
	\centering
	\begin{tabular}{c|c|c}
		\toprule
		$b/T^2_H$ & n=1 & n=2\\ \hline
		0 & $\pm 3.119452-2.746676i$ & $\pm 5.169521-4.763570i$\\
		2.48 & $\pm 3.143638-2.761258i$ & $\pm 5.192327-4.808130i$\\
		5.17 & $\pm 3.211341-2.803189i$ & $\pm 5.246431-4.930215i$\\
		8.44 & $\pm 3.319144-2.873182i$ & $\pm 5.309472-5.114889i$\\
		12.95 & $\pm 3.474557-2.980537i$ & $\pm 5.376798-5.362916i$\\
		20.17 & $\pm 3.706464-3.154053i$ & $\pm 5.468700-5.739936i$\\
		34.84	& $\pm 4.105908-3.490995i$ & $\pm 5.556184-6.661389i$\\
		\bottomrule
	\end{tabular}
	\caption{Lowest two QNM frequencies (scalar sector) ($k=0$) for magnetic black brane}
	\label{tab:Tb4}
\end{table}

\subsubsection{QNMs of Near-Extremal Magnetic Black Branes}

Plot (Fig. 7) shows behavior of imaginary part of lowest QNM frequency for tensorial sector as the value of magnetic field is varied $b/T^2_H \approx 1.5\times 10^3$ to $3.8\times 10^3$. It is a log-log plot with $\log(\left|\mathrm{Im}[\omega-\omega_0]\right|)$ on the \textit{y}-axis and $\log(\pi T_H/b^{1/2})$ on the \textit{x}-axis represented by the blue dots. $\omega_0$ is the lowest QNM fequency computed for the highest value of $b/T^2_H\approx 3.8\times 10^3$ before computational errors get too large to yield meaningful results. The red line shows a linear fit with slope 1 and this is observed for all modes. Unlike the Reisnner-Nordstrom case, the real part $\mathrm{Re}[\omega-\omega_0]$ does not show such scaling with $\pi T_H/b^{1/2}$.

\begin{figure} 
	\centering
	\includegraphics[width=0.6\textwidth]{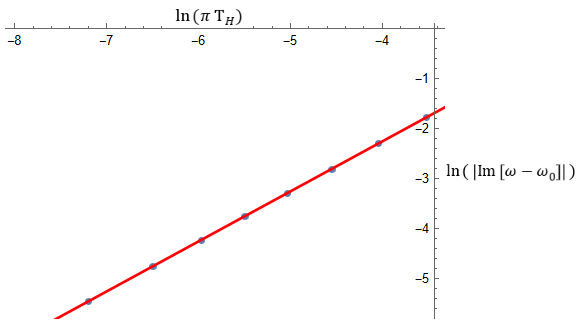}
	\caption{Plot of $\log(\left|\mathrm{Im}[\omega-\omega_0]\right|)$ vs. $\log(\pi T_H/b^{1/2})$ for $b/T^2_H \approx 1.5\times 10^3$ to $3.8\times 10^3$. The red line shows a linear fit $y=x+c_I$}
	\label{fig:Fig7}
\end{figure}

Thus, a linear scaling between $\mathrm{Im}[\omega^{(n)}]$ and $\pi T_H/b^{1/2}$ is seen close to extremality.
\begin{equation}
	\mathrm{Im}\left[\omega^{(n)}\right] \approx \mathrm{Im}\left[\omega^{(n)}_0\right] + \gamma_n\cdot \left(\frac{\pi T_H}{b^{1/2}}\right), \qquad\qquad \left(\frac{T_H}{b^{1/2}} \ll 1\right)
	\label{eq:Eq65}
\end{equation}

\noindent where $n$ is the mode \# and $\gamma_n$ are mode specific constants. From (\ref{eq:Eq65}), decay times for tensorial perturbations (proportional to $\mathrm{Im}[\omega^{(n)}]$) of near-extremal magnetic black branes only depend on their Hawking temperature (in units of magnetic field) $T_H/b^{1/2}$. Table. \ref{tab:Tb6} lists the values of $\gamma_n$ for the lowest ten QNM frequencies. 
\begin{table}[h]
	\centering
	\begin{tabular}{c|c}
		\toprule
		n & $\gamma_n$\\ \hline
		1 & $-5.627282$\\
		2 & $-11.338257$\\
		3 & $-15.788416$\\
		4 & $-19.799393$\\
		5 & $-24.860070$\\
		6 & $-30.175361$\\
		7 & $-34.071120$\\
		8 & $-37.788665$\\
		9 & $-43.785514$\\
		10 & $-48.479225$\\
		\bottomrule
	\end{tabular}
	\caption{$\gamma_n$ for lowest ten tensorial QNM frequencies ($k=0$) of magnetic black branes}
	\label{tab:Tb6}
\end{table}

\section{Conclusion}

The study of quasinormal modes in AdS\textsubscript{5} gravity duals of $\mathcal{N}= 4$ SYM plasmas provides important information on transport coefficients and other dynamic properties otherwise not easily computable in lattice-gauge models. Literature, especially on QNMs of magnetically charged black branes, is scarce and does not have a high amount of detail. In this article, we employ a pseudospectral approach using Chebyshev polynomial basis to solve Einstein field equations for perturbations of asymptotically AdS\textsubscript{5} spacetimes. In particular, we compute the QNMs for tensor metric perturbations of Reisnner-Nordstrom black branes as well as tensor and scalar metric perturbations of magnetic black branes in the limit of vanishing spatial momenta. For both cases, we find that the imaginary part of tensorial sector (non-hydrodynamic) QNM frequencies, and therefore perturbation decay times, display a linear scaling with temperature as extremality is approached which is only known to be true in general for hydrodynamic QNMs.

\newpage
\printbibliography 

@BOOK{Boyd,
	title = {{C}hebyshev and {F}ourier spectral methods},
	publisher = {Courier Corporation},
	author = {Boyd, J.~P.},
	year = {2001},
	edition = {},
}

@BOOK{Sol:2014,
	title = {Gauge/string duality, hot QCD and heavy ion collisions},
	publisher = {Cambridge University Press},
	author = {Casalderrey-Solana, J. and Liu, H. and Mateos, D. and Rajagopal, K. and Wiedemann, U.~A.},
	year = {2014},
	edition = {},
}

@ARTICLE{Ar:2014,
	author = {Aref'eva, I.~Y.},
	title = {{H}olographic approach to quark-gluon plamsa in heavy ion collisions},
	journal = {Physics-Uspekhi},
	year = {2014},
	volume = {57},
	pages = {},
	number = {6},
	month = {},
	publisher = {},
	doi = {},
}

@ARTICLE{Yaf:2015,
	author = {Fuini, J.~F. and Yaffe, L.~G.},
	title = {{F}ar-from-equilibrium dynamics of a strongly coupled non-Abelian plasma with non-zero charge density or external magnetic field},
	journal = {Journal of High Energy Physics},
	year = {2015},
	volume = {2015},
	pages = {1-48},
	number = {7},
	month = {},
	publisher = {},
	doi = {},
}

@ARTICLE{Yaf:2014,
	author = {Chesler, P.~M. and Yaffe, L.~G.},
	title = {{N}umerical solution of gravitational dynamics in asymptotically anti-de Sitter spacetimes},
	journal = {Journal of High Energy Physics},
	year = {2014},
	volume = {2014},
	pages = {1-68},
	number = {7},
	month = {},
	publisher = {},
	doi = {},
}

@ARTICLE{Kam:2016,
	author = {Janiszewski, S. and Kaminski, M.},
	title = {Quasinormal modes of magnetic and electric black branes versus far from equilibrium anisotropic fluids},
	journal = {Physical Review D},
	year = {2016},
	volume = {93},
	pages = {},
	number = {2},
	month = {},
	publisher = {},
	doi = {},
}

@ARTICLE{Nak:2006,
	author = {Nakamura, S. and Sin, S.~J.},
	title = {A holographic dual of hydrodynamics},
	journal = {Journal of High Energy Physics},
	year = {2006},
	volume = {2006},
	pages = {},
	number = {9},
	month = {},
	publisher = {},
	doi = {},
}

@ARTICLE{Mal:2000,
	author = {Aharony, O. and Gubser, S. and Maldacena, J. and Ooguri, H. and Oz, Y.},
	title = {Large N field theories, string theory and gravity},
	journal = {Physics Reports},
	year = {2000},
	volume = {323},
	pages = {183-386},
	number = {3},
	month = {},
	publisher = {},
	doi = {},
}

@ARTICLE{Star:2009,
	author = {Berti, E. and Cardoso, V. and Starinets, A.~O.},
	title = {Quasinormal modes of black holes and black branes},
	journal = {Classical and Quantum Gravity},
	year = {2009},
	volume = {26},
	pages = {},
	number = {16},
	month = {},
	publisher = {},
	doi = {},
}


\end{document}